\documentclass[conference]{IEEEtran}
\IEEEoverridecommandlockouts
\usepackage{cite}
\usepackage{amsmath,amssymb,amsfonts}
\usepackage{algorithmic}
\usepackage{graphicx}
\usepackage{textcomp}
\usepackage{xcolor}
\usepackage[hyphens]{url}
\def\BibTeX{{\rm B\kern-.05em{\sc i\kern-.025em b}\kern-.08em
    T\kern-.1667em\lower.7ex\hbox{E}\kern-.125emX}}
\usepackage{flushend}

\begin{document}

\title{Support and Scandals in GameFi dApps: \\A Network Analysis of \emph{The Sandbox} Transactions\\
%\thanks{Identify applicable funding agency here. If none, delete this.} % TODO: Put Algorand Foundation Here
}

\author{
\IEEEauthorblockN{Fernando Spadea}
\IEEEauthorblockA{Rensselaer Polytechnic Institute\\
Troy, NY \\
% 110 8th St, Troy, NY, USA\\
spadef@rpi.edu\\
% Orcid: 0009-0006-4278-3666
}
\and
\IEEEauthorblockN{Oshani Seneviratne}
\IEEEauthorblockA{Rensselaer Polytechnic Institute\\
Troy, NY \\
% 110 8th St, Troy, NY, USA\\
senevo@rpi.edu\\
% Orcid: 0000-0001-8518-917X
}
}

\maketitle

\begin{abstract}
We explore the burgeoning field of GameFi through a detailed network analysis of \emph{The Sandbox}, a prominent decentralized application (dApp) in this domain. Utilizing the bow-tie model, we map out transaction data within \emph{The Sandbox}, providing a novel perspective on its operational dynamics. Our study delves into the varying impacts of external support, uncovering a surprising absence of enduring effects on network activity. We also investigate the network’s response to several notable incidents, including the Ronin Hack and the United States Securities and Exchanges Commission's hearing on cryptocurrencies, revealing a generally resilient structure with limited long-term disturbances. A critical aspect of our analysis focuses on the `whales,' or major stakeholders in \emph{The Sandbox}, where we uncover their pivotal role in influencing network trends, noting a significant shift in their engagement over time. This research sheds light on the intricate workings of GameFi ecosystems and contributes to the broader discourse on the intersection of the Web, AI, and society, particularly in understanding the resilience and dynamics of emerging digital economies. We particularly note the parallels of the long-tail behavior we see in web-based ecosystems appearing in this niche domain of GameFi. Our findings hold significant implications for the future development of equitable and sustainable GameFi dApps, offering insights into stakeholder behavior and network resilience in the face of external challenges and opportunities.
\end{abstract}


\section{Introduction}
Our research explores several factors contributing to the market uncertainty found in GameFi dApps. We analyze \emph{The Sandbox}, a popular GameFi dApp, by studying the effects of support from traditional brands on its activity and the effects of various scandals. On top of that, we determine the types of users contributing to \emph{The Sandbox}'s short-term and long-term success. We leverage the bow-tie model to map out transaction data within \emph{The Sandbox}, offering a novel perspective on its operational dynamics and underlying graph structure. Our study delves into the varying impacts of external support, uncovering a surprising absence of enduring effects on network activity. We also investigate the network's response to several notable incidents, including the Ronin Hack and the United States Securities and Exchanges Commission's hearing on cryptocurrencies, revealing a generally resilient structure with limited long-term disturbances. Although GameFi dApps are similar to traditional web-based games in many ways, they differ in that they have whales or major stakeholders. A critical aspect of our analysis focuses on the whales, where we uncover their pivotal role in influencing network trends. We note a significant shift in their engagement over time, highlighting their importance in understanding the short-term success of GameFi platforms. However, we also find that a dApp's long-term success relies heavily on building a dedicated user base.

\section{Background}

With the rise of blockchain technology and decentralized applications (dApp), we have seen the rise and fall of decentralized finance (DeFi) over the past few years. During that time, we have seen the volatile nature of society's perception of said technology. Because these DeFi dApps were largely not based on any tangible value, this perception played a large role in the success and failure of DeFi. Many events, both positive and negative, helped shape said perception and thus affected how much people interacted with DeFi dApps. Therefore, analyzing these events would reveal the resilience of DeFi markets and different strategies for increasing or maintaining activity.

However, alongside DeFi, GameFi faced the same dilemma. As the name implies, GameFi is similar to DeFi but has a gaming element. Games are built as dApps that use a blockchain to manage in-game asset ownership. Gamification in the blockchain space has been highly legitimized recently, even being used by the Andiami project to fix the problem of centralization in decentralized blockchains \cite{prnewswireAndiamiUnveiled}. One of the most successful GameFi dApps was Axie Infinity\footnote{\url{https://axieinfinity.com/}} in which players battled with monsters called Axies. A player's in-game items and monsters were tracked on the Ronin blockchain as NFTs. The in-game currency also took the form of a token. This gives players complete control over their game assets, opening up the possibility of selling those assets to other players or investors to make a profit. Thus, these GameFi games enable players to earn value by playing them. This is known as the play-to-earn model, and it proved to be lucrative for many players worldwide for a while, and many even started playing these games full-time \cite{axieAppointment}. However, the GameFi market crashed along with DeFi, leaving many of these players out of a job \cite{axieDisappointment}. On top of this, Axie Infinity later suffered a massive hack named the Ronin hack, which further crashed the GameFi market \cite{cointelegraphAftermathAxie}. This reflected a shift in public opinion on blockchain in general. However, a separate GameFi dApp existed by the name of \emph{The Sandbox}\footnote{\url{https://www.sandbox.game/}}.

\subsection{\emph{The Sandbox}}
\emph{The Sandbox} is a unique GameFi dApp because it is not so much a game as a marketplace and game engine for developers to create games and assets for players to interact with \cite{sandboxWhitepaper}. It is more so a competitor to Roblox\footnote{\url{https://www.roblox.com/}} than to any traditional game. \emph{The Sandbox} is hosted on the Ethereum blockchain. In \emph{The Sandbox}, developers can buy LAND NFTs representing plots of land in the game's map that the players can explore. \emph{The Sandbox}'s world is designed and populated by its LAND owners, so they have to host games and events on said LAND to attract players \cite{sandboxWhitepaper}. By doing this, these developers can profit by monetizing their visitors via entrance fees or possibly selling them in-game items. Other developers can also profit by creating ASSETs, assets that can be used in games hosted by LAND owners \cite{sandboxWhitepaper}. ASSETs can also target players, such as clothing or equipment. All of this is tied together with the SAND token, which is the currency that is used by players and developers alike when purchasing goods and services in \emph{The Sandbox} \cite{sandboxWhitepaper}. This creates a marketplace with many different types of users that interact with each other. It is not just the LAND owners that can make money in \emph{The Sandbox}, but also any developers willing to put time into creating ASSETs targeted at players and LAND owners. As a result, \emph{The Sandbox} has a unique marketplace with many different types of users, making it a more notable topic of study than other types of GameFi dApps.

\section{The Data}

\subsection{Transaction Graph}
We have gathered transaction data from Etherscan for our analysis of \emph{The Sandbox}. This transaction data consists of all transactions involving any of the contracts created by \emph{The Sandbox} Deployer\footnote{0x18dd4e0eb8699eA4FeE238dE41ECfb95e32272f8} and \emph{The Sandbox} Deployer 2\footnote{0xe19ae8F9B36Ca43D12741288D0e311396140DF6F} addresses on the Ethereum blockchain. This includes normal, token, NFT, multi-token, and internal transactions as defined by Ethereum and Etherscan. This is to say that we have collected every transaction related to \emph{The Sandbox}. Therefore, this includes any transactions involving SAND, LAND, or ASSETs. The data spans from October 10, 2019, to October 26, 2023, so it covers most of \emph{The Sandbox}'s lifespan up to this point. In total, we have 4,948,956 transactions involving 659,248 addresses. We use the \textit{from}, \textit{to}, \textit{contractAddress}, \textit{timestamp}, and \textit{value} parameters of the data collected. The \textit{from} parameter holds the address that sent the transaction, while the \textit{to} parameter holds the address the transaction was targeted at. Sometimes, the \textit{to} parameter is empty, so, in those cases, we use the \textit{contractAddress} parameter instead, which holds the address of the contract invoked in the transaction. The \textit{timestamp} parameter holds when the transaction was sent, and the \textit{value} parameter holds the amount of Ethereum sent in the transaction.

To find potential vulnerabilities within \emph{The Sandbox}'s codebase, we also gathered GitHub issue data on all of \emph{The Sandbox}'s repositories. However, there are only 14 with a collective two issues, which do not raise any concerns with \emph{The Sandbox}'s code. One asks a question, and the other answers a question found in the comments of one of the repository's files. \emph{The Sandbox} also scored a high AA on their CertiK audit \cite{certikSandboxCertiK}, so it is fairly safe to say that \emph{The Sandbox}'s code does not have many if any, security risks by current standards. Thus, this data bears no fruit.

\subsection{Bow-tie Model}
To start, we create a directed network from our transaction data where the nodes represent addresses, and the edges are the transactions from and to each other. Then, we split this network into the Bow-tie Model to further analyze its underlying graph structure \cite{glattfelder2019bow}. The Bow-tie model comprises several groups: SCC, IN, OUT, TUBES, TENDRILS IN, TENDRILS OUT, and OTHER. The SCC, the strongly connected component, is the collection of nodes in the graph that interact with each other. The IN group is the collection of nodes that interact with the SCC but do not have the SCC interact with them. The OUT group consists of the nodes that are interacted with by the SCC but do not interact with the SCC. The TUBES are a series of nodes that connect the IN group to the OUT group. The TENDRILS IN are nodes that are interacted with by the IN group but are disconnected from the SCC and OUT groups. The TENDRILS OUT are nodes that interact with the OUT group but are disconnected from the SCC and IN groups. The OTHER nodes are just those that do not fit into any of these categories \cite{glattfelder2019bow}.

Separating the network into these partitions shows the differences between their activity levels at any given time. This helps us identify how these different groups reacted to different events and whether those reactions were positive or negative. We will also be able to better analyze the graph's structure as we will have similarly behaving addresses grouped for closer analysis.

\subsection{\emph{The Sandbox} Support}
\label{sec:dataSupp}
A point of interest for \emph{The Sandbox} is the outside support it receives from traditional brands. The brands from which \emph{The Sandbox} has received support vary greatly in their fields. Some of the support has come from traditional video game companies such as Square Enix with Dungeon Siege \cite{mediumSandboxSquare}, Ubisoft with Rabbids \cite{mediumSandboxUbisoft}, Skybound Entertainment with The Walking Dead (also a comic book and TV show) \cite{animocabrandsSandboxAnnounces}, Atari \cite{venturebeatAtariLaunches}, and ZEPETO \cite{mediumZEPETOPartners}. They have also received support from several music artists such as Snoop Dogg \cite{mediumSnoopDogg}, deadmau5 \cite{mediumDeadmau5Sandbox}, Steve Aoki \cite{mediumSteveAokis}, and the Warner Music Group \cite{wmgSandboxPartners}. Adidas \cite{venturebeatAdidasOriginals} and Gucci Vault \cite{forbesGucciVault} also provided support as fashion brands, and the Smurfs \cite{animocabrands} and Care Bears \cite{mediumCareBears} were present as well as Cartoon Brands. We chose these brands as our points of interest as these are the brands that \emph{The Sandbox} consistently points to when discussing companies that are invested in \emph{The Sandbox}.

\subsection{\emph{The Sandbox} Scandals}
\label{sec:dataScand}
On the other end of support are the scandals that \emph{The Sandbox} has faced. As discussed in the data section, \emph{The Sandbox} has a relatively good record compared to its competitors, so it lacks scandals related to \emph{The Sandbox} code itself. But there are a few related scandals we will analyze. The first scandal we looked at was the Ronin hack mentioned above. While this hack did not directly involve \emph{The Sandbox}, it did affect the value of SAND \cite{coingape622Million}, so it is worth investigating the activity level after it occurred. The next scandal we looked at involved an employee's work computer being hacked and used to send phishing emails to Sandbox users whose emails the computer had access to \cite{dailycoinHackersTarget}. While this event did not involve a vulnerability in \emph{The Sandbox}'s code, it reflected poorly on the company's security standards so that it could have alerted some users to action. Another similar scandal involved the CEO's Twitter being hacked and used to post a crypto scam \cite{cointelegraphSandboxCEOs}. Again, this did not directly involve the security of \emph{The Sandbox}, but it reflects poorly upon its parent company. The last scandal we looked at was the recent United States Securities and Exchanges Commission (SEC) hearing where SAND was labeled an unregistered security \cite{secSECgovCharges}. While \emph{The Sandbox} has claimed that they disagree with this claim \cite{decryptSandboxCOO}, this is still a significant event that affected the outlook of GameFi as a whole, so looking at how this affected the transaction data is worthwhile.

\subsection{Whales}
\label{sec:whales}

With our transaction data, we also look at the SAND whales of \emph{The Sandbox}. Whales, in the context of cryptocurrency, are addresses that hold a substantial amount of cryptocurrency. This paper defines a whale as an address that has received a net amount of SAND worth at least 1e22 Wei or 10,000 Eth. By identifying these whales, we get insight into the number of whales involved in \emph{The Sandbox} and their reactions to these events. We also identify the change in whales over time and which sections of the bow-tie model they fit into.

\section{Methodology}

As discussed in sections \ref{sec:dataSupp} and \ref{sec:dataScand}, \emph{The Sandbox} has received support from several traditional brands and has faced several scandals over time, but the impact of these events on \emph{The Sandbox}'s success is unclear. Thus, we aim to analyze this impact using our transaction data. By looking at the dates these collaborations were released, we see if they resulted in increased or decreased transactions. Specifically, we look at the 30 days before and after the events to identify any spikes and changes in patterns surrounding the events. We also looked at the total value of these transactions in the same 61-day period to see if these events affected the volume of value sent. We also use our bow-tie model to analyze the effects of these events on the different partitions we made to see how the different groups react to each event. With these separate groups, we can see how these events affected activity among different types of users in \emph{The Sandbox} and if there were any significant shifts in the graph's structure. We analyze the shifting structure of the transaction graph by calculating the bow-tie partitions before and after these events to see how they affected their sizes and, thus, the graph's structure. We also specifically look at where these moving addresses are coming from, whether they are newly introduced to the graph or shifted from another partition.

With the whales discussed in section \ref{sec:whales}, we perform a similar analysis with the Bow-tie model and these events to see how the overall patterns of the network compare to the patterns of just the transactions involving the whales. Also, by looking at the changing amount of SAND held by different addresses over time, we identified if there were whales that bailed on SAND and thus stopped being whales and whether their transaction patterns revealed anything about the graph. Also, as mentioned previously, we identify which categories within the Bow-tie model these whales exist within and discuss the implications of those classifications. In addition, we use the average degree of the network and the whales as a point of analysis.

\section{Results}

\subsection{Bow-tie Model}
Our complete network has a total of $659,248$ nodes. When we split the network into the components of the bow-tie model, we find that most of the network falls into the SCC and OUT partitions. The SCC accounts for $390,531$ nodes ($\sim59.25\%$) while the OUT accounts for $264,501$ nodes ($\sim40.13\%)$. The IN only accounts for $4,071$ nodes ($\sim0.62\%$). The IN TENDRILS, OUT TENDRILS, and OTHER groups account for 7, 136, and 2 nodes respectively. We found the TUBES partition to be empty.

\begin{figure}[htbp!]
    \centering
    \includegraphics[width=\columnwidth]{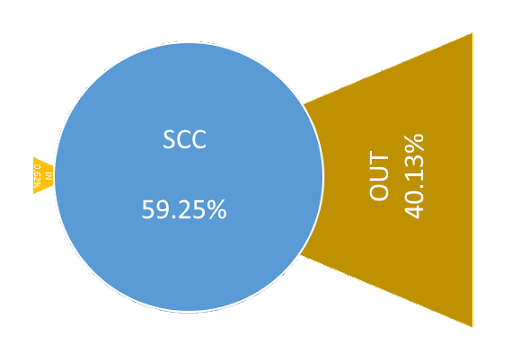}
    \caption{Diagram of our Network in the Bow-tie Model}
    \label{fig:Bow-tieDiagram}
    %\Description[Diagram of our network in the Bow-tie Model]{A diagram showing how our network fits into the Bow-tie model. The SCC, IN, and OUT groups account for $59.25\%$, $40.13\%$, and $0.62\%$ of the nodes, respectively.}
\end{figure}

Since most user transactions in \emph{The Sandbox} involve some transaction in response, we don't see many addresses in the IN category. It mostly consists of inactive or bot addresses and addresses approving the use of the SAND token for the trade-in of another contract. On the other hand, the OUT group holds more investors than normal users. One would have to interact with \emph{The Sandbox} to get anything from its game aspect, so this makes sense. Then, in the SCC, we have the bulk of it making up \emph{The Sandbox}'s gaming community. These primary users drive the value of SAND by actually using it to play in \emph{The Sandbox}.

\subsection{\emph{The Sandbox} Support}

\begin{figure*}[htbp]
    \centering
    \includegraphics[width=\textwidth]{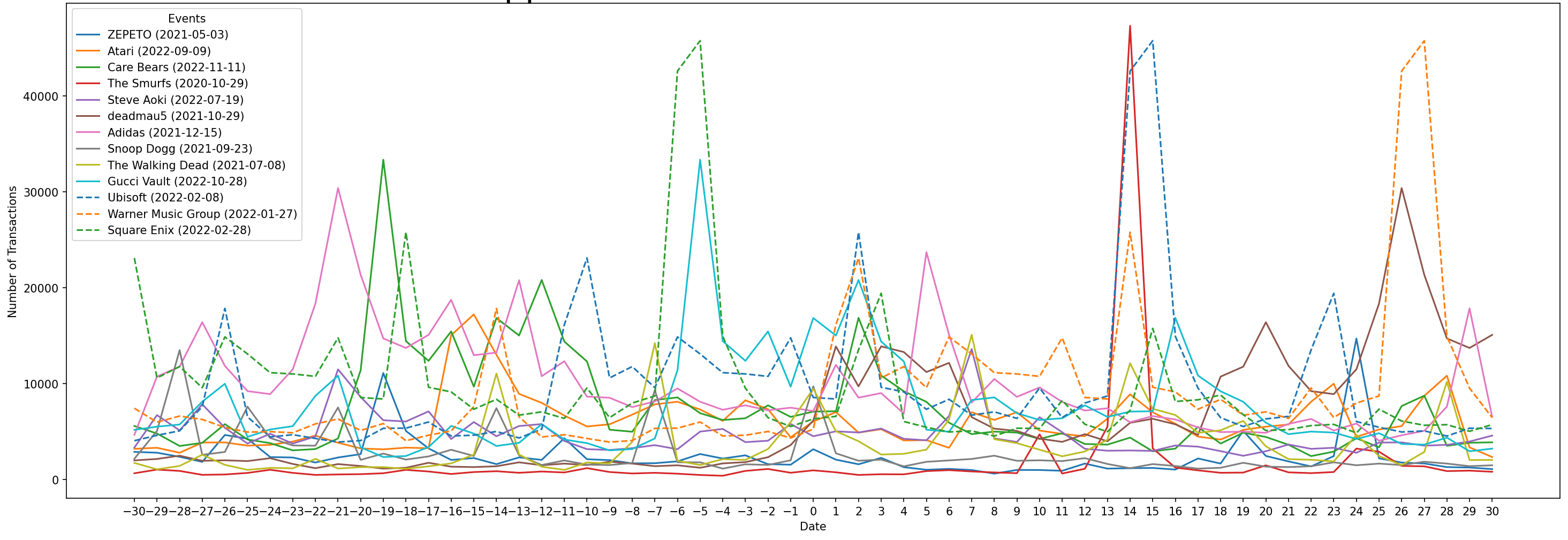}
    \caption{Graph showing the number of daily transactions in the 30 days before and after each support event explored.}
    \label{fig:CondensedSupportNumber}
    %\Description[Number of Transactions vs. Date]{The support events shown are ZEPETO on May 3, 2021, Atari on September 9, 2022, Care Bears on November 11, 2022, The Smurfs on October 29, 2020, Steve Aoki on July 19, 2022, deadmau5 on October 29, 2021, Adidas on December 15, 2021, Snoop Dogg on September 23, 2021, The Walking Dead on July 8, 2021, Gucci Vault on October 28 2022, Ubisoft on February 8 2022, Warner Music Group on January 27 2022, and Square Enix February 28 2022.}
\end{figure*}

\begin{figure*}[htbp]
    \centering
    \includegraphics[width=\textwidth]{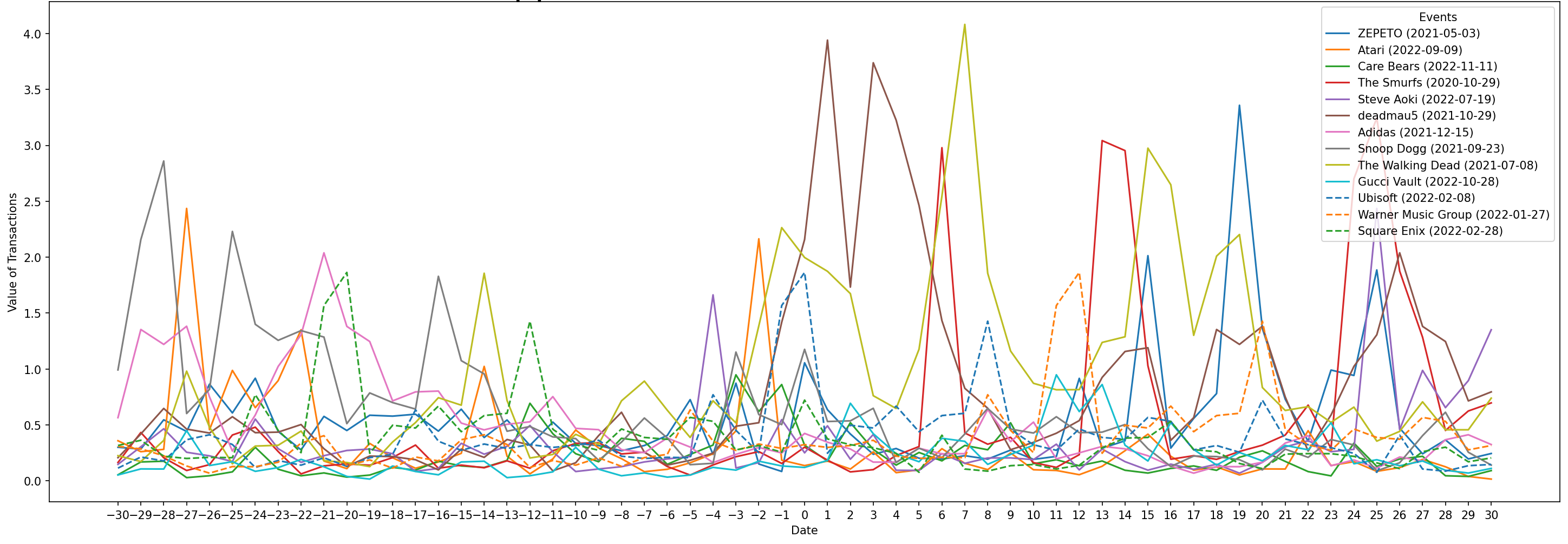}
    \caption{Graph showing the total value of transactions daily in the 30 days before and after each support event explored.}
    \label{fig:CondensedSupportValue}
    %\Description[Value of Transactions vs. Date]{The support events shown are ZEPETO on May 3, 2021, Atari on September 9, 2022, Care Bears on November 11, 2022, The Smurfs on October 29, 2020, Steve Aoki on July 19, 2022, deadmau5 on October 29, 2021, Adidas on December 15, 2021, Snoop Dogg on September 23, 2021, The Walking Dead on July 8, 2021, Gucci Vault on October 28 2022, Ubisoft on February 8 2022, Warner Music Group on January 27 2022, and Square Enix February 28 2022.}
\end{figure*}

After calculating the number of transactions per day for the entire range of our dataset, we plotted the values for the 30 days before and after the dates attributed to each ``support event'' in figure \ref{fig:CondensedSupportNumber}. We did the same for the total value of the transactions each day in figure \ref{fig:CondensedSupportValue}. Both graphs show short-lived spikes in transactions and value but no long-lasting effects after these spikes stabilize. The two exceptions to this are deadmau5 and Warner Music Group. However, the consistent increase in daily interactions after deadmau5's support is likely attributed to the simultaneous GameFi boom. On the other hand, Warner Music Group's effect is not so easily written off. Nevertheless, the rest of the support events quickly returned to the norm, meaning they failed to attract a consistent user base.

\begin{figure}[htbp]
    \centering
    \includegraphics[width=\columnwidth]{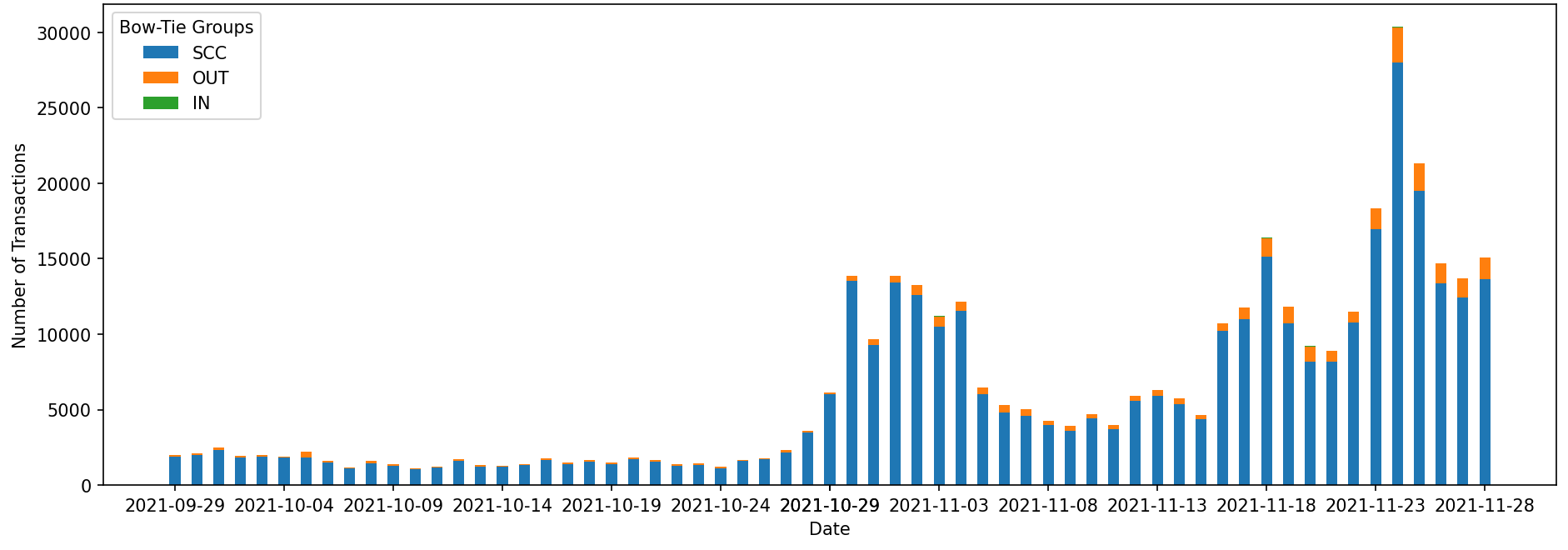}
    \caption{Stacked bar graph showing the number of transactions per day in the 30 days before and after October 29, 2021, when deadmau5's project was released. The transactions are split into three groups depending on whether they are part of the SCC, OUT, or IN groups of the bow-tie model.}
    \label{fig:stackedDeadmau5}
    %\Description[Number of Transactions vs. Date]{Stacked bar graph showing the number of transactions per day in the 30 days before and after October 29, 2021, when deadmau5's project was released. The transactions are split into three groups depending on whether they are part of the SCC, OUT, or IN groups of the bow-tie model.}
\end{figure}

\begin{figure}[htbp]
    \centering
    \includegraphics[width=\columnwidth]{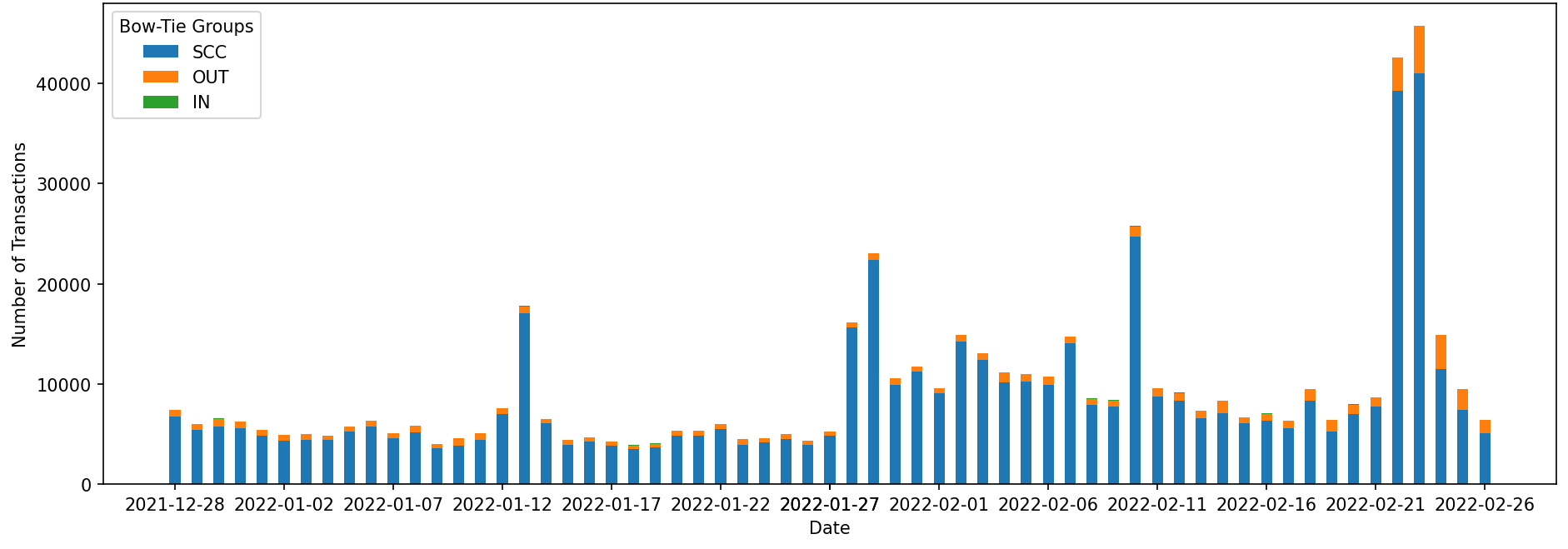}
    \caption{Stacked bar graph showing the number of daily transactions in the 30 days before and after January 27, 2022, when Warner Music Group's project was released. The transactions are split into three groups depending on whether they are part of the SCC, OUT, or IN groups of the bow-tie model.}
    \label{fig:stackedWarner}
    %\Description[Number of Transactions vs. Date]{Stacked bar graph showing the number of transactions per day in the 30 days before and after January 27, 2022, when Warner Music Group's project was released. The transactions are split into three groups depending on whether they are part of the SCC, OUT, or IN groups of the bow-tie model.}
\end{figure}

\begin{figure}[htbp]
    \centering
    \includegraphics[width=\columnwidth]{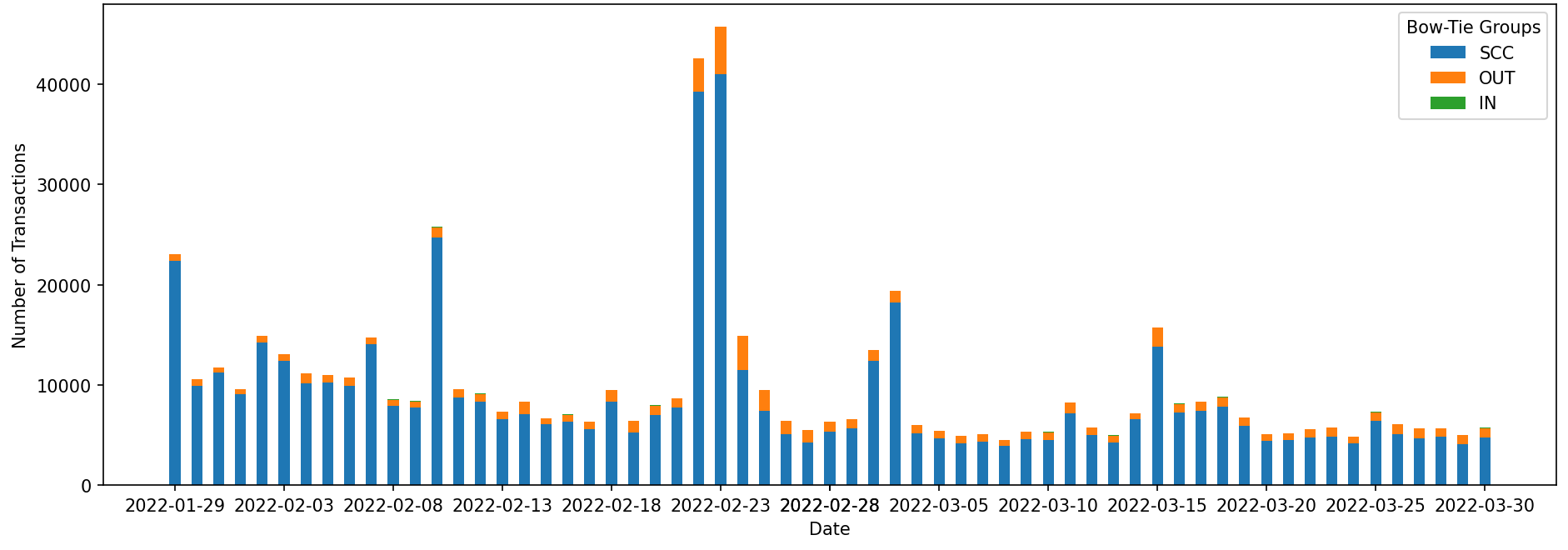}
    \caption{Stacked bar graph showing the number of daily transactions in the 30 days before and after February 28, 2022, when Square Enix's project was released. The transactions are split into three groups depending on whether they are part of the SCC, OUT, or IN groups of the bow-tie model.}
    \label{fig:stackedSquare}
    %\Description[Number of Transactions vs. Date]{Stacked bar graph showing the number of transactions per day in the 30 days before and after February 28, 2022, when Square Enix's project was released. The transactions are split into three groups depending on whether they are part of the SCC, OUT, or IN groups of the bow-tie model.}
\end{figure}

To take a closer look at these results, we have graphed the data for deadmau5, Warner Music Group, and Square Enix in figures \ref{fig:stackedDeadmau5}, \ref{fig:stackedWarner}, and \ref{fig:stackedSquare}, respectively. Here, we have split the transactions based on whether they belong to the bow-tie model's SCC, OUT, or IN groups.

In the deadmau5 graph, figure \ref{fig:stackedDeadmau5}, there is an initial spike in SCC transactions followed by a dip that stays well above the number of transactions each day throughout the previous month. This indicates a reliable increase in activity. There has also been a noticeable increase in the number of transactions belonging to the OUT group. It is not as significant as the SCC's but still significant relative to its previous values. This indicates an increase in investors. The IN group does not have a noticeable change in activity, which could be attributed to its small size. However, we do have to keep in mind that this support event coincided with the widespread boom of the GameFi market.

There is a similar pattern in the Warner Music Group's graph, figure \ref{fig:stackedWarner}. There is an increase in activity in the SCC and the OUT groups without anything noticeable in the IN group. However, the increase in OUT transactions is less noticeable since the previous month's values are higher than they were with deadmau5. Regardless, this again indicates an increase in the active user base and investors. Additionally, this effect cannot be as easily attributed to another event, so it is likely due to the Warner Music Group's efforts in \emph{The Sandbox}. 

We use Square Enix's support to represent the rest of the support events, as they all had similarly unspectacular results. In the Square Enix graph, figure \ref{fig:stackedSquare}, there is a very short-lived spike in SCC transactions followed by an immediate return to normalcy. The OUT and IN transactions are not noticeably affected. The other support events saw similar results\footnote{More data and graphs showing these results can be generated or found in the GitHub repository linked in the Resource Contributions.} except for The Smurfs, which had no effect, likely due to how early into \emph{The Sandbox}'s life they supported it. Therefore, Warner Music Group is the exception rather than the rule, and most of these events did not have a long-lasting effect on the activity in \emph{The Sandbox}.

\subsection{\emph{The Sandbox} Scandals}

\begin{figure*}[htbp]
    \centering
    \includegraphics[width=\textwidth]{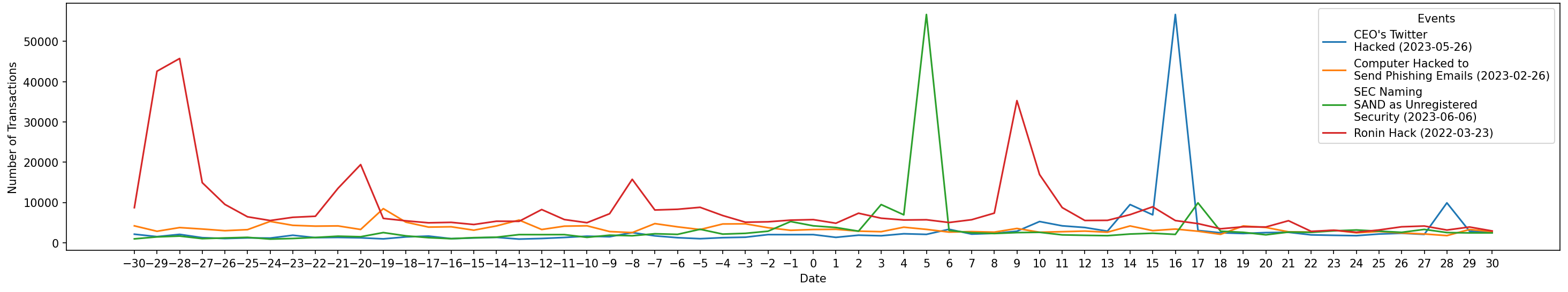}
    \caption{Graph showing the number of transactions per day in the 30 days before and after each of the scandals explored.}
    \label{fig:CondensedScandalNumber}
    %\Description[Number of Transactions vs. Date]{The scandals shown are \emph{The Sandbox}'s CEO's Twitter account being hacked and used to post a crypto scam on May 26, 2023, a Sandbox employee's computer being hacked and used to send phishing emails to Sandbox users on February 26, 2023, the SEC naming SAND as an unregistered security on June 6, 2023, and the Ronin Hack on March 23, 2022.}
\end{figure*}

\begin{figure*}[htbp]
    \centering
    \includegraphics[width=\textwidth]{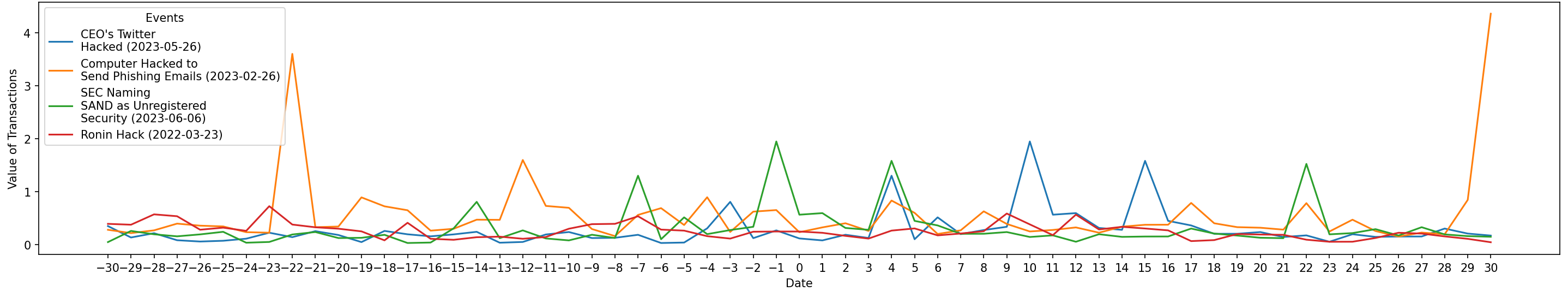}
    \caption{Graph showing the total value of transactions each day in the 30 days before and after each of the scandals explored.}
    \label{fig:CondensedScandalValue}
    %\Description[Value of Transactions vs. Date]{The scandals shown are \emph{The Sandbox}'s CEO's Twitter account being hacked and used to post a crypto scam on May 26, 2023, a Sandbox employee's computer being hacked and used to send phishing emails to Sandbox users on February 26 2023, the SEC naming SAND as an unregistered security on June 6, 2023, and the Ronin Hack on March 23, 2022.}
\end{figure*}

We created similar graphs of the number of transactions and the total value of transactions for our scandals in figures  \ref{fig:CondensedScandalNumber} and \ref{fig:CondensedScandalValue}. In both graphs, we do not see any long-lasting effects due to the CEO's Twitter account being hacked or the Employee's computer being hacked and used to send phishing emails. In the value graph, the same applies to the SEC naming SAND unregistered security and the Ronin Hack. However, the effects of these scandals are more noticeable in the transactions graph, with the Ronin Hack having long-term effects.

\begin{figure}[htbp]
    \centering
    \includegraphics[width=\columnwidth]{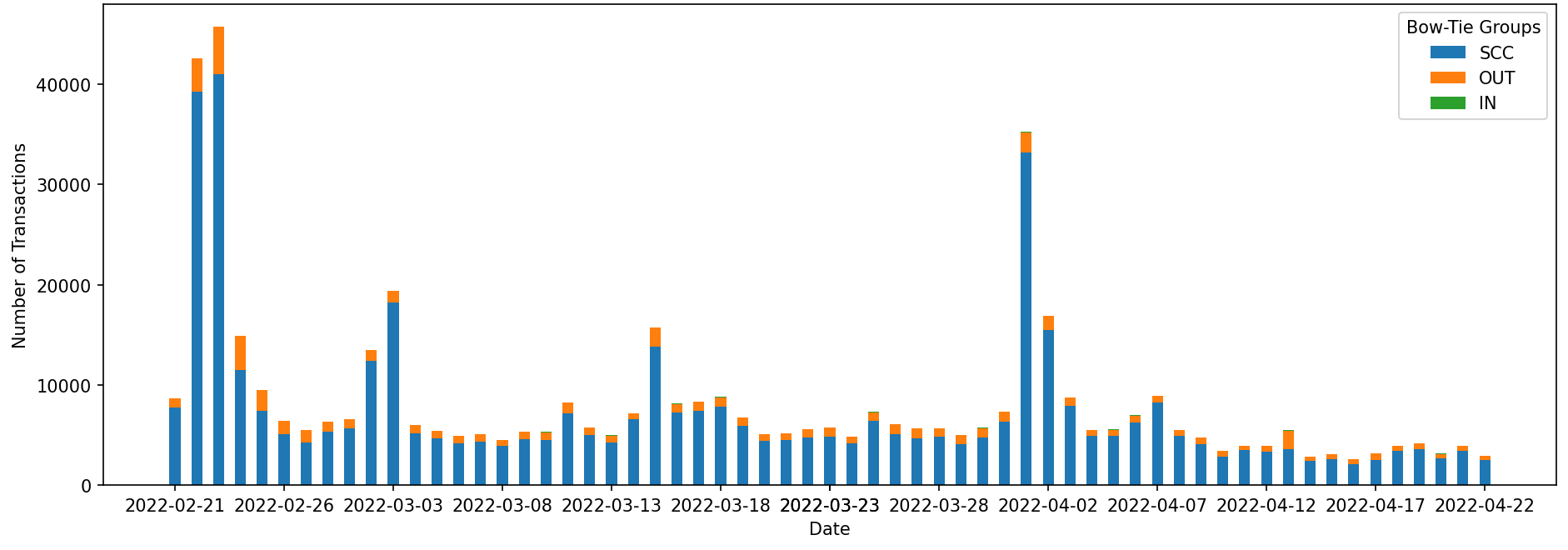}
    \caption{Stacked bar graph showing the number of transactions per day in the 30 days before and after March 23, 2022, when the Ronin Hack occurred. The transactions are split into three groups depending on whether they are part of the SCC, OUT, or IN groups of the bow-tie model.}
    \label{fig:stackedRonin}
    %\Description[Number of Transactions vs. Date]{Stacked bar graph showing the number of transactions per day in the 30 days before and after March 23, 2022, when the Ronin Hack occurred. The transactions are split into three groups depending on whether they are part of the SCC, OUT, or IN groups of the bow-tie model.}
\end{figure}

\begin{figure}[htbp]
    \centering
    \includegraphics[width=\columnwidth]{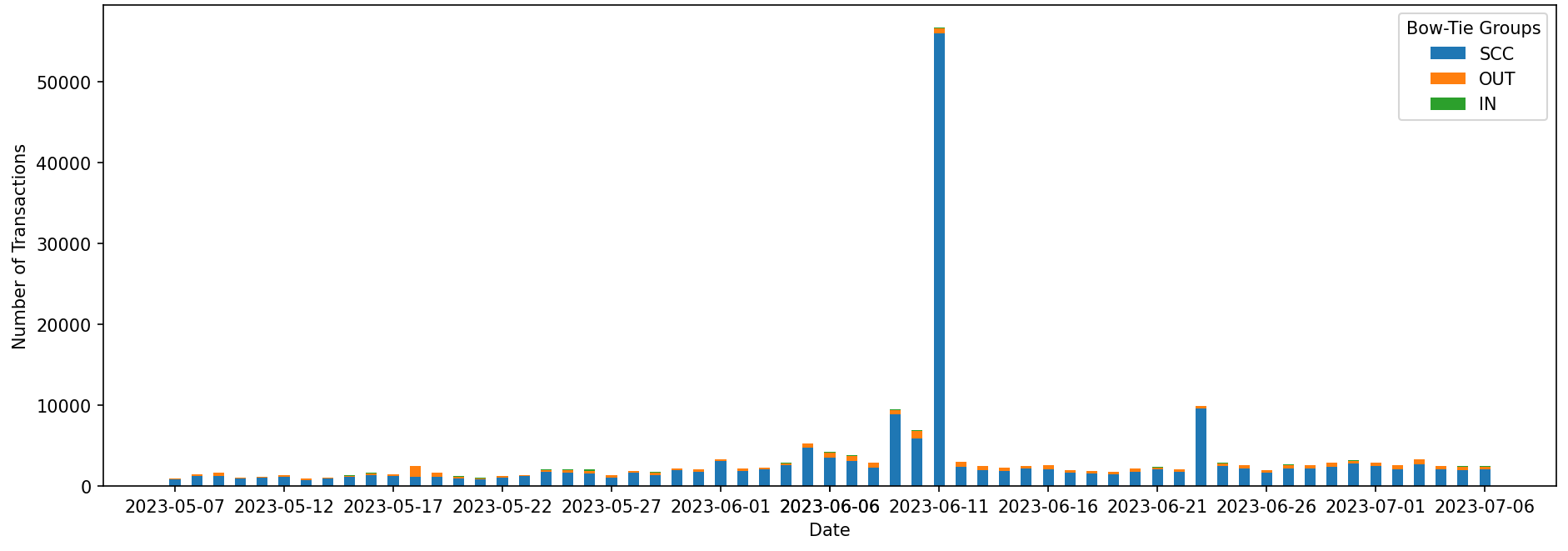}
    \caption{Stacked bar graph showing the number of daily transactions in the 30 days before and after June 6, 2023, when the SEC named SAND an unregistered security. The transactions are split into three groups depending on whether they are part of the SCC, OUT, or IN groups of the bow-tie model.}
    \label{fig:stackedSEC}
    %\Description[Number of Transactions vs. Date]{Stacked bar graph showing the number of transactions per day in the 30 days before and after June 6, 2023, when the SEC named SAND an unregistered security. The transactions are split into three groups depending on whether they are part of the SCC, OUT, or IN groups of the bow-tie model.}
\end{figure}

\begin{figure}[htbp]
    \centering
    \includegraphics[width=\columnwidth]{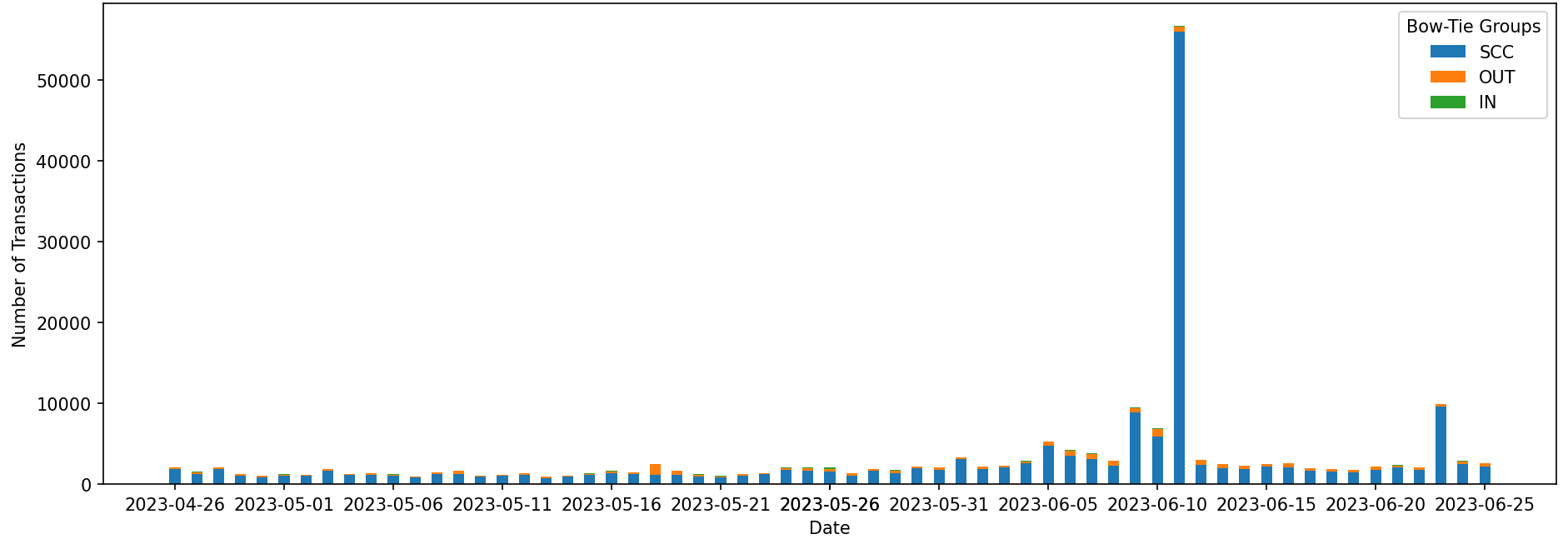}
    \caption{Stacked bar graph showing the number of transactions per day in the 30 days before and after May 26, 2023, when the Twitter account of the  CEO of \emph{The Sandbox} was hacked and used to post a crypto scam. The transactions are split into three groups depending on whether they are part of the SCC, OUT, or IN groups of the bow-tie model.}
    \label{fig:stackedCEO}
    %\Description[Number of Transactions vs Date]{Stacked bar graph showing the number of transactions per day in the 30 days before and after May 26, 2023, when the Twitter account of the  CEO of \emph{The Sandbox} was hacked and used to post a crypto scam. The transactions are split into three groups depending on whether they are part of the SCC, OUT, or IN groups of the bow-tie model.}
\end{figure}

Similarly to what we did for the support events, we have graphed the data for the Ronin Hack, SEC naming SAND a security, and the CEO's Twitter account being hacked in figures \ref{fig:stackedRonin}, \ref{fig:stackedSEC}, and \ref{fig:stackedCEO}, respectively. Once again, we split the transactions based on whether they belong to the bow-tie model's SCC, OUT, or IN groups.

In the Ronin Hack graph, figure \ref{fig:stackedRonin}, there is a delayed spike in SCC and OUT transactions followed by a dip that drops slightly below the previous days. The reason for the delay is that the Ronin hack took about a week to be noticed. This result indicates that the initial spike was due to users and investors jumping ship since there are fewer daily transactions after the spike. The long-lasting effect of this scandal is clear despite \emph{The Sandbox} not being directly affected by the Ronin Hack.

In the SEC graph, figure \ref{fig:stackedSEC}, there is a massive spike in SCC transactions a few days after the event, but there is not much of a change in the number of daily transactions afterward. This may seem odd, but it is worth noting that this event happened far after \emph{The Sandbox}'s prime, so its user base was barely active. This is further evidenced by the lack of a reaction in the OUT group, which indicates that most of the investors in the OUT group who would have been scared off by this event had already left.

We use the CEO's Twitter hacking as a representative for itself and the Employee computer hacking since they had similar results\footnote{More data and graphs showing these results can be generated or found in the GitHub repository linked in the Resource Contributions.}. In figure \ref{fig:stackedCEO}, the hack has no noticeable effect on the activity in \emph{The Sandbox}. Like with the SEC event, it is important to consider that this scandal occurred well after the prime of \emph{The Sandbox}, so the user base was already at its weakest point, with the users left not being shaken by these sorts of scandals.

\subsection{Change in Bow-Tie Model Over Time}

\begin{figure*}[htbp]
    \centering
    \includegraphics[width=.91\textwidth]{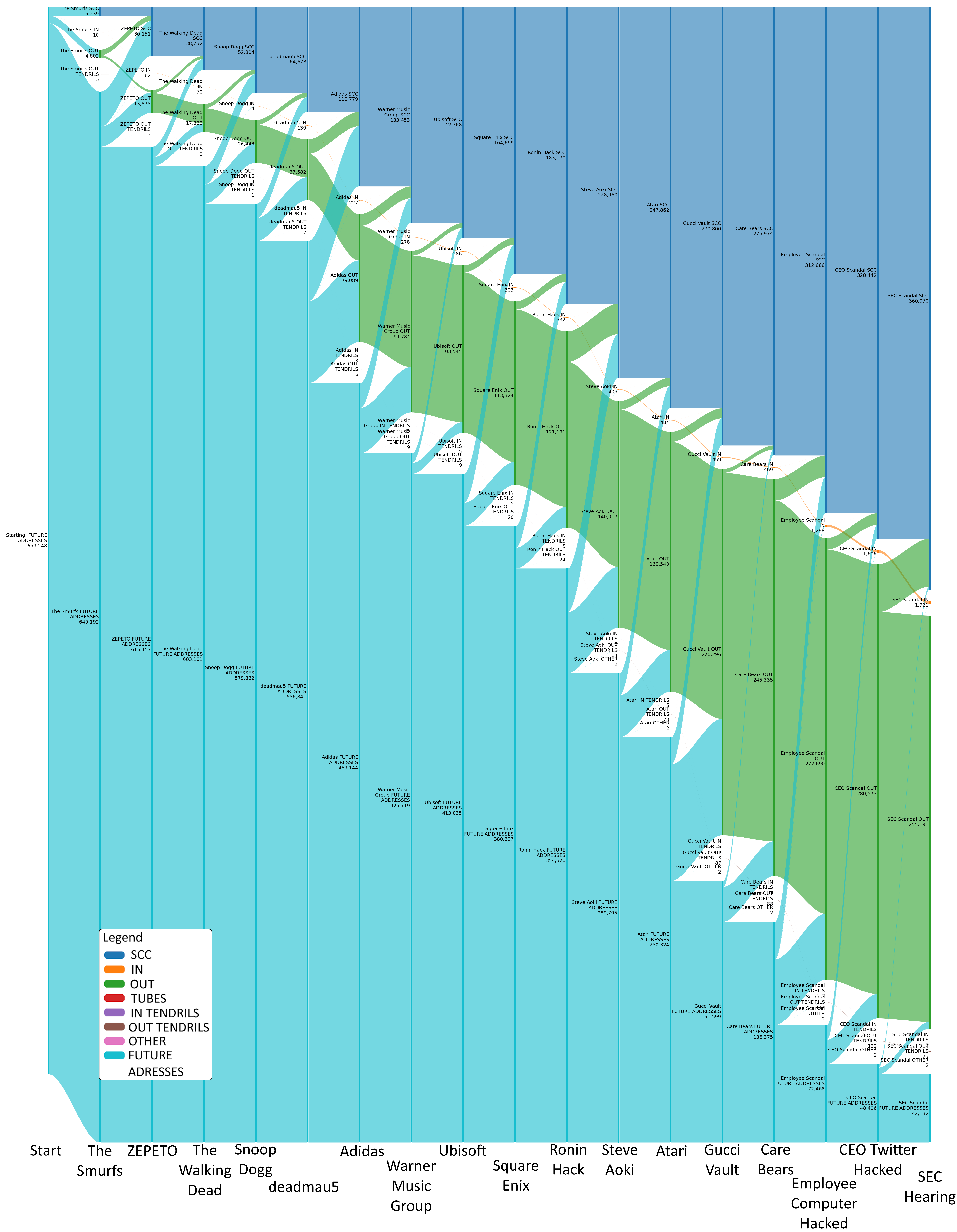}
    \caption{Sankey graph showing the change in the Bow-Tie model of the network over time ten days after each of the events explored, combining both the support events and scandals.}
    \label{fig:sankey}
    %\Description[Bow-Tie Model Change Over Time Sankey Graph]{The support events shown are ZEPETO on May 3, 2021, Atari on September 9, 2022, Care Bears on November 11, 2022, The Smurfs on October 29, 2020, Steve Aoki on July 19, 2022, deadmau5 on October 29, 2021, Adidas on December 15, 2021, Snoop Dogg on September 23, 2021, The Walking Dead on July 8, 2021, Gucci Vault on October 28, 2022, Ubisoft on February 8, 2022, Warner Music Group on January 27, 2022, and Square Enix February 28 2022. The scandals shown are \emph{The Sandbox}'s CEO's Twitter account being hacked and used to post a crypto scam on May 26, 2023, a Sandbox employee's computer being hacked and used to send phishing emails to Sandbox users on February 26, 2023, the SEC naming SAND as an unregistered security on June 6, 2023, and the Ronin Hack on March 23 2022.}
\end{figure*}

In figure \ref{fig:sankey}, we graph the changes in the categories of the bow-tie model over time in our Sandbox network. We also include future addresses not yet involved in the network to illustrate when new addresses will be introduced. Note that addresses cannot leave the SCC once they have entered it. The groups other than the SCC and OUT are small throughout the graph.

The SCC grows steadily between events, but we see three significant jumps after Adidas, Steve Aoki, and the SEC hearing. The Adidas jump can likely be attributed to the GameFi boom. The Steve Aoki jump from the OUT group could, in part, be attributed to the lasting effects of the Ronin hack, causing members of the OUT group to sell their investments to members of the SCC. In contrast, the new addresses that entered the SCC could be attributed to new users taking advantage of the dropping value of SAND. The influx of addresses from the OUT to the SCC after the SEC hearing is likely a result of investors selling their investments, often to members of the SCC.

The growth of the OUT is less consistent than that of the SCC, partly because the OUT can lose members while the SCC cannot. Nevertheless, there are several significant increases in new OUT addresses. The first is after Adidas, which, like with the SCC, can be attributed to the GameFi boom and, thus, a significant increase in investors. We see a decent increase in new OUT addresses after the Warner Music Group support, which matches what we saw in our closer analysis of this event in figure \ref{fig:stackedWarner}, where we saw a slight increase in OUT activity. We see a similar pattern between the Ronin Hack and Steve Aoki as we did with the SCC, which once again indicates that investors may have been hoping the effect of the Ronin Hack was just a dip that would be recovered from\footnote{\url{https://cryptoquant.com/asset/sand/chart/market-data/price-volume?market=spot&exchange=all_exchange&window=DAY&sma=0&ema=0&metricScale=linear}}. After the Gucci Vault event, there is a surprisingly significant increase in the OUT group. Although this does match a spike in OUT transactions before the Gucci Vault event present in the data\footnote{This data and graphs are available in the GitHub repository in Resource Contributions.}, this cannot be attributed to the Gucci Vault event since it occurred afterward. There is also another surprising increase in the OUT between the Care Bears event and the Employee's computer being hacked. This is likely due to a slight increase in value that SAND saw between these two events\footnote{\url{https://cryptoquant.com/asset/sand/chart/market-data/price-volume?market=spot&exchange=all_exchange&window=DAY&sma=0&ema=0&metricScale=linear}} which attracted investors again. It is difficult to attribute this increase to either of these two events since they did not cause any significant increases in OUT group activity\footnote{This data and graphs are available in the GitHub repository in Resource Contributions.}.

Overall, the SCC's growth is more consistent than the surprisingly unpredictable OUT. It is generally difficult to attribute the significant increases in addresses to the events explored in this paper since the increases generally do not match the increases in activity that resulted from these events except for the Warner Music Group support.

\subsection{Whales}

\begin{figure*}[htbp]
    \centering
    \includegraphics[width=\textwidth]{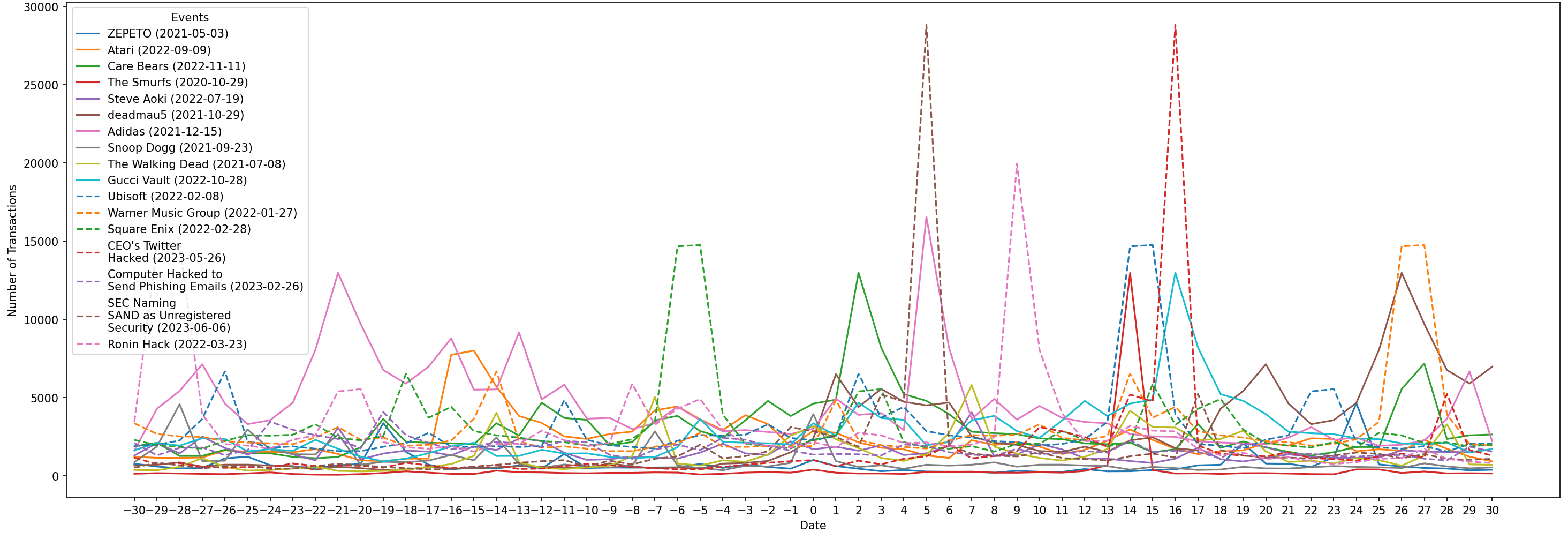}
    \caption{Graph showing the number of daily transactions involving current Sandbox whales in the 30 days before and after each event explored.}
    \label{fig:CondensedWhaleNumber}
    %\Description[Number of Transactions vs. Date]{The support events shown are ZEPETO on May 3, 2021, Atari on September 9, 2022, Care Bears on November 11, 2022, The Smurfs on October 29, 2020, Steve Aoki on July 19, 2022, deadmau5 on October 29, 2021, Adidas on December 15, 2021, Snoop Dogg on September 23, 2021, The Walking Dead on July 8, 2021, Gucci Vault on October 28, 2022, Ubisoft on February 8 2022, Warner Music Group on January 27 2022, and Square Enix February 28 2022. The scandals shown are \emph{The Sandbox}'s CEO's Twitter account being hacked and used to post a crypto scam on May 26, 2023, a Sandbox employee's computer being hacked and used to send phishing emails to Sandbox users on February 26, 2023, the SEC naming SAND as an unregistered security on June 6, 2023, and the Ronin Hack on March 23, 2022.}
\end{figure*}

\begin{figure*}[htbp]
    \centering
    \includegraphics[width=\textwidth]{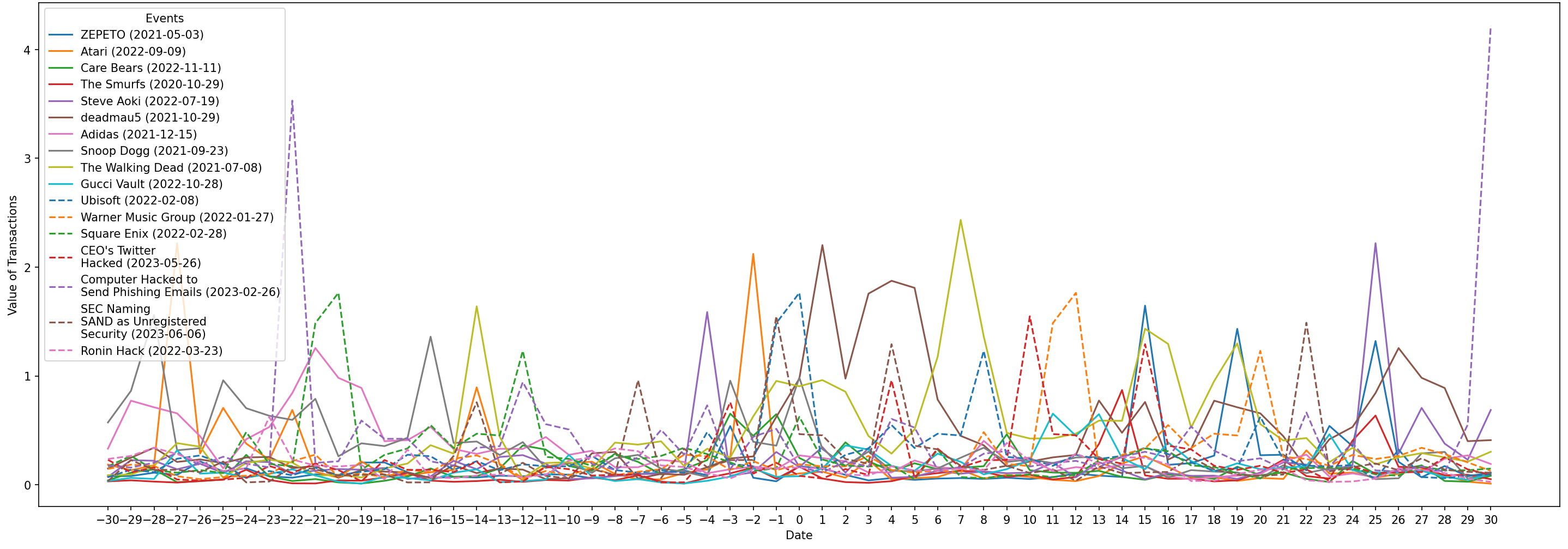}
    \caption{Graph showing the total value of daily transactions involving current Sandbox whales in the 30 days before and after each event explored.}
    \label{fig:CondensedWhaleValue}
    %\Description[Value of Transactions vs. Date]{The support events shown are ZEPETO on May 3, 2021, Atari on September 9, 2022, Care Bears on November 11, 2022, The Smurfs on October 29, 2020, Steve Aoki on July 19, 2022, deadmau5 on October 29, 2021, Adidas on December 15, 2021, Snoop Dogg on September 23, 2021, The Walking Dead on July 8, 2021, Gucci Vault on October 28, 2022, Ubisoft on February 8 2022, Warner Music Group on January 27 2022, and Square Enix February 28 2022. The scandals shown are \emph{The Sandbox}'s CEO's Twitter account being hacked and used to post a crypto scam on May 26, 2023, a Sandbox employee's computer being hacked and used to send phishing emails to Sandbox users on February 26, 2023, the SEC naming SAND as an unregistered security on June 6, 2023, and the Ronin Hack on March 23, 2022.}
\end{figure*}

We found 2,464 whales invested in \emph{The Sandbox} as of October 26, 2023. Of these whales, 890 are in the SCC, and 1,574 are in the OUT group. However, we also check to see if any addresses ever met this requirement at some point during the three years our data covers. There were 33,758 qualifying whales over time, with 32,174 in the SCC and 1,584 in the OUT group. As we can see, most of the former whales were in the SCC, which makes sense since the SCC would be the most likely group of nodes to offload one's SAND on. We combined the transaction and value graphs for the support events and scandals in figures  \ref{fig:CondensedWhaleNumber} and \ref{fig:CondensedWhaleValue}, respectively. We combined the graphs because they show almost identical patterns to the normal graphs, so the same analysis applies. This implies that whales are largely involved in most transactions in \emph{The Sandbox}, which indicates that some of these whales could function as resources for \emph{The Sandbox}. To avoid filling this paper with more graphs, we can see how these patterns will also hold when we include the past whales since they will just be adding more transactions to a pattern that already matches the pattern created by the entire network\footnote{All the code to generate these graphs and the graphs themselves are included in the GitHub repository linked in the Resource Contributions.}. This is further supported by the average degree of the network compared to the average degree of the whales. The average degree of the network is about 12.866, 5.395 when we exclude all whales, while the average degree of whales is 735.520, 151.286 if we include past whales. This demonstrates that whales interact with significantly more addresses than normal users.

\section{Related Work}

The bow-tie model has previously been used to analyze the Steemit blockchain \cite{guidi2020steem}, where they found that the SCC made up a comfortable majority of the transaction network at 56.3\%, similar to our 59.25\%. The OUT group was the second largest at 43.7\%, which is also similar to our 40.13\%. They found that the third largest group was the IN, with a significantly lower number of nodes than in the SCC and OUT groups at 0.016\%, like with our 0.62\% IN group, and the rest of the groups were insignificantly small. Overall, the groups of their bow-tie model were very similar to ours, so this could be a common pattern seen in more blockchains. They also analyzed the whales of the Steemit chain and found that they held a significant amount of power, which could also be a common pattern. It is clear that this analysis strategy has its merits and could be applied to several different blockchain dApps, as we have been able to apply similar analysis strategies to \emph{The Sandbox}. We have also possibly found patterns that could be universal across blockchains, so more research could be done to see if this pattern holds.

% In another paper on the Steemit blockchain, 
Li et al. use the Steemit blockchain transaction graph to measure how decentralized it is and analyze its bot activity \cite{li2019incentivized}. Their results reveal that a few powerful addresses in reality controlled the network, similar to our analysis of whales in \emph{The Sandbox}. Additionally, they also found significant bot activity abusing the reward system on Steemit. While this analysis differs from our paper's approach, it shows the significant power that a blockchain's transaction graph holds for analysis.

Guidi et al. analyzes Decentraland, another popular GameFi project where users can buy land in the form of NFTs \cite{guidi2022social}. Here, they analyze how the NFT market affects the in-game world and how users interact with the game itself. They find that the profitability of the platform works against its playfulness. Once again, this paper analyzes a different aspect of a GameFi project, showing much potential for analysis in these projects.

In \cite{lai2023quantitative}, Lai et al. analyze the transaction graph of Axie Infinity to discover player behavior patterns and measure the effect of whales on the game. They found that, while many users were actively playing the game, there was a worrying amount of hoarding of game assets by a small number of addresses, which they believed could threaten the game's future. This is similar to our analysis of whales in \emph{The Sandbox} and their significant influence, and we additionally once again see the power of the transaction graph.

\section{Conclusion}
In the end, we saw that most support from outside parties received by \emph{The Sandbox} did not significantly impact its long-term activity. The only exceptions found were Warner Music Group and deadmau5, with deadmau5's supposed impact likely attributable to the general rise of GameFi that coincides with their support. This means that these attempts to attract users via popular traditional brands did not result in any shift in public perception that legitimized \emph{The Sandbox} enough to attract a new wave of consistent activity. DApps hoping to achieve this will have to look to other strategies. This approach may support this end, but it needs to be part of a larger plan.

The scandals looked at also had largely insignificant effects on the long-term activity of \emph{The Sandbox} other than the Ronin hack, which shook the entire GameFi space. However, this could be because the other three scandals occurred after \emph{The Sandbox}'s prime when its activity was already very low. This indicates that the users still involved in \emph{The Sandbox} at the time of these scandals were more invested and less likely to be shaken by these scandal levels. Thus, these users form \emph{The Sandbox}'s true core user base that will survive these perturbations.
Our analysis showed that whales are involved in most transactions in \emph{The Sandbox} environment and largely drive the overall activity. We also saw that many whales dropped \emph{The Sandbox} and stopped being whales. However, we found that these whales were similarly involved to the surviving whales regarding \emph{The Sandbox}'s activity. This driving force that whales have is significant since it means that they are largely responsible for the activity on \emph{The Sandbox}, so it could be beneficial to target whales directly when trying to drive up activity in GameFi dApps.

Ultimately, GameFi markets, like DeFi, are very unstable and difficult to control. Trying to stabilize those markets positively through third-party support and events is a poor strategy. Viable strategies should instead target whales directly as they are a small subset of the user base with substantial sway over the market and, thus, have much more potential to produce significant results. On the other hand, once a GameFi market has stabilized to a lower state, it is significantly more resilient to both positive and negative events. It remains stable, meaning a small but reliable user base has formed. In combination, this means that whales need to be targeted during unstable periods, but an invested user base needs to be built up over time for the future as those users will stick with the game past its prime. Building a dedicated user base is something that traditional web-based games have excelled at, so GameFi has a great legacy from which to build. Following this approach, GameFi dApps can achieve early high success followed by a lower but stable success in the long term.

% \smallskip
% \noindent\textbf{Resource Contributions:}
\section{Resource Contributions}

Our research artifacts (code, documentation, and graphs) are shared under the Apache 2.0 license. We maintain an open-source GitHub repository for all our artifacts: \url{https://github.com/decentralized-social-media/gameFi}.

% \noindent\textbf{Acknowledgements:}
\section{Acknowledgements}
This work was partially supported by the Algorand Centres of Excellence program managed by the Algorand Foundation. The opinions expressed in this publication do not necessarily represent the views of the Algorand Foundation.

\bibliographystyle{IEEEtran}
\bibliography{sources}

\end{document}